\documentclass[floatfix,twocolumn,prl,longbibliography]{revtex4-1}
\usepackage{graphicx}
\usepackage{dcolumn}
\usepackage{bm}
\usepackage{amsmath}
\usepackage{times}
\usepackage{color}
\usepackage[breaklinks=true,colorlinks,citecolor=blue,linkcolor=blue,urlcolor=blue]{hyperref}
\usepackage{gensymb}

\makeatletter

\newcommand{\Rmnum}[1]{\expandafter\@slowromancap\romannumeral #1@}
\makeatother

\begin{document}
\title{Chiral orbital current driven topological Hall effect in Mn$_3$Si$_2$Te$_6$}
\author{Arnab Das}
\affiliation{Department of Physics, Indian Institute of Technology Kanpur, Kanpur 208016, India}

\author{Soumik Mukhopadhyay}
\email{soumikm@iitk.ac.in}
\affiliation{Department of Physics, Indian Institute of Technology Kanpur, Kanpur 208016, India}  
    
\begin{abstract}
    Chiral orbital current (COC) plays a crucial role in governing the magnetization and transport behaviour in the layered ferrimagnetic nodal-line semiconductor Mn$_3$Si$_2$Te$_6$. Here, we observe that the topological Hall effect (THE), typically attributed to Berry curvature from chiral spin textures, originates from COC, which produces an emergent magnetic field for conduction electrons due to its real-space orbital textures. We find that the THE signal strengthens as we move down from bulk to nanoflakes, but tends to disappear with increasing current, along with the disappearance of the COC state. We also demonstrate a strong correlation between the colossal magnetoresistance (CMR) and the observed THE, suggesting that large Berry curvature and topological transport can arise purely from orbital degrees of freedom, providing a new platform for engineering dissipationless transport in 2D magnets.
\end{abstract}

\maketitle

{\it Introduction:---}
The interplay of topology, magnetism, and orbital degrees of freedom has emerged as a central theme in condensed matter physics, giving rise to a variety of unconventional transport phenomena. In particular, the topological Hall effect (THE) is widely regarded as a transport fingerprint of nontrivial spin chirality and Berry curvature associated with real-space spin textures. With the discovery of the skyrmion lattice in the chiral ferromagnet MnSi \cite{Muhlbauer2009}, THE has emerged as a platform to explore exotic physical properties. In recent times, THE has been observed in many non-centrosymmetric systems \cite{Neubauer2009,Kanazawa2011,Surgers2014,Gallagher2017}, and in a few centrosymmetric systems as well \cite{Wang2016,Zheng2021,Rout2019,Kurumaji2019}. In most systems, THE is shown to originate from broken inversion symmetry giving rise to the Dzyaloshinsky-Moriya interaction (DMI) which favours the formation of topologically nontrivial spin textures like skyrmions, anti-skyrmions, skyrmonium, (anti)merons, bimerons, etc \cite{Nayak2017,Zhang2018,Gao2020,Yu2018,Nagase2021, Muhlbauer2009}. Recent theoretical developments, however, suggest that orbital degrees of freedom alone can generate topological–chiral interactions through emergent orbital magnetism associated with chiral orbital currents (COC), independent of conventional spin–orbit coupling (SOC) \cite{PhysRevB.98.195439,Grytsiuk2020}. Such orbital-driven interactions have been predicted to stabilize exotic magnetic textures and to engender new routes toward topological transport responses \cite{Grytsiuk2020,Yang2023MonopoleOrbital,Go2018Intrinsic,Nikolaev2024ChiralOrbital,dosSantosDias2016ChiralityDrivenOrbital,PhysRevResearch.5.043052}.

COC has been under investigation since the early 2000s with studies on high-T$_\mathrm{C}$ cuprates \cite{Varma2006, Varma2014, Bourges2021, DiMatteo2012, Pershoguba2013, Scagnoli2011, Pershoguba2014, Yakovenko2015}, iridates \cite{Zhao2015,Jeong2017,Murayama2021}, Kagome superconductors and magnets \cite{Jiang2021,Feng2021,Teng2022} and moir\'e
heterostructures \cite{Serlin2020,Tschirhart2021,Sharpe2019}. Later it was also shown that the coupling of COC with long-range magnetic order can break both mirror and time-reversal symmetry and can lead to interesting phenomena like colossal magnetoresistance (CMR) \cite{Zhang2022ChiralOrbitalCurrents}. Coupling the COC with applied current can also present intriguing physics, especially giving rise to a current sensitive unique Hall effect in Mn$_3$Si$_2$Te$_6$ (MST) \cite{Zhang2024CurrentSensitiveHall}. In MST, the COC flows along the Te edges in the ab-plane, giving rise to an orbital moment $\mathrm{M_{COC}}$ and a real-space emergent magnetic field B$_\mathrm{c}$, both directed along the c-axis. Currently, most of the studies on MST are directed towards the unique phenomena of CMR \cite{Seo2021ColossalAMR,PhysRevB.111.174419,PhysRevB.106.045106,PhysRevB.103.L161105,Zhang2024,Lovesey2023} with very few findings on the Hall resistivity of the same. The observation of the unique the Hall effect in MST cannot be completely explained away by considering the contributions of ordinary Hall effect (OHE) and anomalous Hall effect (AHE) \cite{Zhang2024CurrentSensitiveHall, PhysRevB.108.125103}. Thus, the complete picture of the exact contributions to the Hall effect of MST and the physics behind it is still unknown. 

In this Letter, we observe that the Hall effect in both bulk and nanoflake devices of MST corresponds very well with its CMR phenomena, indicating a common origin. The temperature and current dependence of the Hall signal reveals a very strong connection between the observed Hall effect and the COC in MST, suggesting that the latter might be the driving force behind the  observed uncoventional Hall effect in MST. 

{\it Experimental Methods:---}
Single crystals of Mn$_3$Si$_2$Te$_6$ (MST) have been prepared using the chemical vapour transport (CVT) method, where iodine (I$_2$) is used as the transport agent (see Sec.~I of SM \cite{SM} for details on sample preparation). The typical crystals were of 1-2 mm in size. Energy dispersive spectroscopy (EDS) measurements were performed to confirm the elemental composition of the grown crystals. X-ray diffraction (XRD) measurements were performed using a PANalytical Empyrean diffractometer. Both powder XRD and single crystal XRD measurements were carried out to confirm the purity and crystallinity of the grown crystals. The powder XRD data was profile fitted using P$\Bar{3}$m space group symmetry using FullProf$\_$Suite software. The temperature dependence of magnetization and the isothermal magnetization hysteresis measurements of the bulk single crystals have been carried out using a Quantum Design physical properties measurement system (PPMS) using the standard vibrating sample magnetometry (VSM) technique. For magento-transport measurements, we use a variable temperature insert (VTI) cryostat provided by CRYOGENIC Inc., UK. The raw MR and Hall curves were symmetrized and antisymmetrized, respectively, to eliminate the effects of electrode misalignment.
\begin{figure}
  \includegraphics[width=0.9\linewidth]{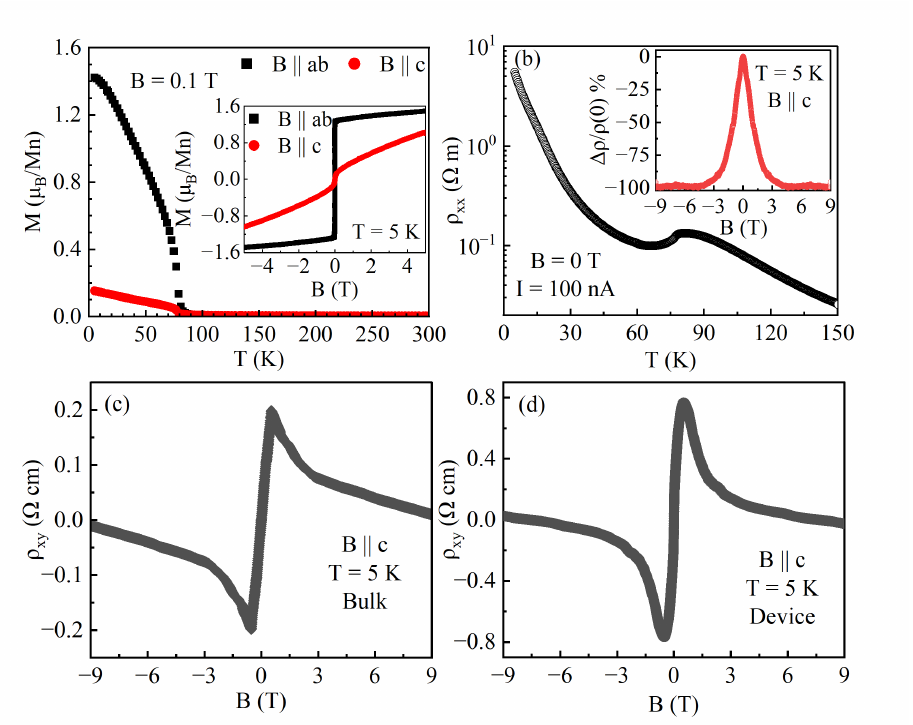}
  \caption{(a) Temperature dependence of magnetization M(T) of bulk MST samples at B = 0.1 T along B $\|$ ab and B $\|$ c directions. Inset: Isothermal magnetization M(B) at T = 5 K for both B $||$ ab and B $||$ c. (b) Temperature dependence of the electrical resistivity of bulk MST. Inset: Magnetic field dependence of MR$\%$ at 5 K with B $\|$ c. (c) Hall resistivity curve at 5 K for bulk MST samples. (d) Hall resistivity curve at 5 K for the nanoflake MST device.}
  \label{fig1}
\end{figure}

{\it Results and Discussion:---}
The single crystal XRD data of the bulk single crystal reveals the (001) orientation of the crystal surface (refer to Sec.~I of SM \cite{SM} for the details). The powder XRD spectra (see Fig.~1(a) of \cite{PhysRevB.111.174419} for details) reveals that the grown single crystals are of high purity. Additionally, the EDS spectra (refer to Sec.~I of SM \cite{SM} for the details) shows prominent peaks of Mn, Si and Te, confirming that the obtained stoichiometry is close to the targeted one. 
\begin{figure*}[htp]
  \includegraphics[width=0.78\linewidth]{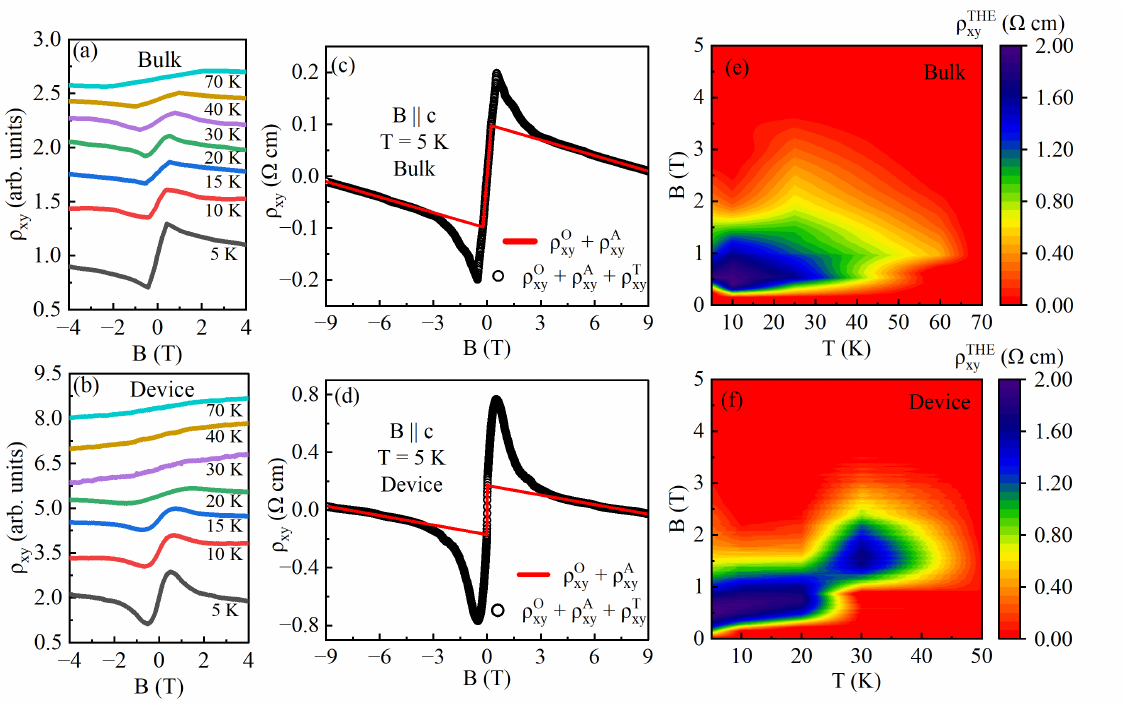}
  \caption{\textcolor{blue}{The field dependence of Hall resistivity ($\mathrm{\rho_{xy}}$) at different temperatures for (a) Bulk and (b) Device of MST, respectively, along B $\|$ c directions. Hall resistivity curves at 5 K, fitted to extract the topological Hall contribution for (c) Bulk and (d) Device of MST, respectively. The black circles represent the raw $\mathrm{\rho_{xy}}$ data, and the red solid line represents the fitted curves. Contour plots of the THE contribution ($\mathrm{\rho_{xy}^{THE}}$) for (e) Bulk and (f) Device of MST, respectively, as a function of temperature and magnetic field.}}
  \label{fig2}
\end{figure*}

Figure~\ref {fig1}(a) shows the temperature-dependent magnetization M(T) of bulk MST single crystals at an applied field of 0.1 T. It shows that the ab-plane is the easy plane where the magnetization saturates at around 1.5 $\mathrm{\mu_B}$/Mn, while the c-axis is the hard axis. The Curie temperature as obtained from Fig.~\ref{fig1}(a) is close to 78 K, which is consistent with earlier reports \cite{PhysRevB.111.174419,PhysRevB.106.045106,PhysRevB.103.245122,PhysRevB.108.125103,PhysRevB.103.L161105,Zhang2024CurrentSensitiveHall,Zhang2022ChiralOrbitalCurrents,Seo2021ColossalAMR,PhysRevB.95.174440}. The inset of Fig.~\ref{fig1}(a) shows the isothermal magnetization at 5 K for both the crystallographic axes, suggesting that the c-axis magnetization does not saturate even at a higher field of 5 T. The temperature dependence of electrical resistivity for the bulk single crystals of MST is shown in Fig.~\ref{fig1}(b), which indicates its semiconducting nature similar to other reports \cite{PhysRevB.106.045106,PhysRevB.103.245122,PhysRevB.108.125103}. The inset of Fig.~\ref{fig1}(b) shows the field dependence of magnetoresistance (MR$\%$) for bulk MST samples at 5 K with B $\|$ c showing the well-known CMR effect. 

Figure~\ref {fig1}(c) shows the field dependence of Hall resistivity ($\mathrm{\rho_{xy}}$) at 5 K for the bulk MST single crystal. The appearance of a distinct sharp peak in the low field region of the Hall resistivity data suggests additional contributions to the $\mathrm{\rho_{xy}}$ data apart from the OHE and AHE, similar to what was observed in previous reports \cite{PhysRevB.108.125103,PhysRevB.103.L161105,Zhang2024CurrentSensitiveHall}. Similar peaks in the field-dependent Hall resistivity of the nanoflake MST device \textcolor{blue}{(Device D1 of SM \cite{SM})} are observed in Fig.~\ref{fig1}(d). \textcolor{blue}{If the observed low-field nonlinearity were dominated by field-induced changes in carrier density or mobility, one would expect a comparable nonlinearity to persist or evolve continuously into the high-field regime. However, it is clear from Fig.~\ref{fig1}(c),(d), $\mathrm{\rho_{xy}}$ depicts a strictly linear field dependence at high fields for both bulk and nanoflake devices. This indicates possible contributions from the THE.}

To further examine the origin of the sharp peak observed in the Hall resistivity data, we measure the field dependence of $\mathrm{\rho_{xy}}$ at different temperatures for both bulk and nanoflake device of MST, as shown in Fig.~\ref{fig2}(a) and \ref{fig2}(b), respectively. \textcolor{blue}{Figures~\ref{fig2}(a),(b) confirms that the peak in $\mathrm{\rho_{xy}}$ remains confined to the same low-field regime over a broad temperature range. The high-field Hall slope also remains linear and unchanged ruling out the possibilty of carrier-based nonlinear Hall mechanisms which typically strengthen at higher fields which is not the case here.} For bulk, the peak in the Hall resistivity data persists up to to 70 K, while for the device, it starts to die out above 30 K. The occurrence of this peak in the $\mathrm{\rho_{xy}}$ curve for bulk MST has been attributed to COC in previous reports \cite{Zhang2024CurrentSensitiveHall}. This COC is known to induce an orbital moment (M$\mathrm{_{COC}}$) along the c axis \cite{Zhang2022ChiralOrbitalCurrents}, that underpins the well-known CMR in bulk MST. This M$\mathrm{_{COC}}$ produces a magnetic field B$\mathrm{_c}$ in the real space, which is directed along the c-axis. When B $\|$ c, it adds up to the external field applied and influences the Hall resistivity \cite{Zhang2024CurrentSensitiveHall}. 

In general, the total Hall resistivity $\mathrm{\rho_{xy}}$ can be expressed as
\begin{eqnarray}
    \mathrm{\rho_{xy}} = \mathrm{\rho_{xy}^O} + \mathrm{\rho_{xy}^A} + \mathrm{\rho_{xy}^T} = \mathrm{R_0B + R_S M_s}+\mathrm{\rho_{xy}^T}
\end{eqnarray}
where $\mathrm{\rho_{xy}^O}$, $\mathrm{\rho_{xy}^A}$ and $\mathrm{\rho_{xy}^T}$, refers to ordinary, anomalous and topological Hall contribution to $\mathrm{\rho_{xy}}$, respectively. M$\mathrm{_s}$ refers to saturation magnetization, while R$_0$ and $\mathrm{R_S}$ are the ordinary and anomalous Hall coefficients, respectively. The ordinary Hall component usually dominates as a high-field linear component of the data and is separated out by calculating R$_0$ using a linear fit at high fields. \textcolor{blue}{For bulk, $\mathrm{R_S}$ is isolated by fitting with the magnetization curve. For the device, it is separated by fitting with a step-like function similar to \cite{wmb35r6b}.} 

\textcolor{blue}{Although in materials exhibiting CMR, R$_0$ and $\mathrm{R_S}$ can be field dependent, however, for MST, the Hall resistivity at sufficiently high magnetic fields is strictly linear, despite the presence of CMR. This allows a robust experimental determination of an effective R$_0$ in the high-field regime where the non-linear Hall effect is completely absent. The anomalous Hall term $\mathrm{R_S}$ depends explicitly on magnetization along B $\|$ c, which increases monotonically with field and does not saturate even at high fields, evident from the inset of Fig.~\ref{fig1}(a). The additional Hall contribution appears only in a narrow low-field window and vanishes well before magnetization saturation, indicating that even if $\mathrm{R_S}$ were allowed to vary smoothly with field, the product $\mathrm{R_S M_s}$ cannot generate a sharp, symmetric peak localized near zero field.} 

Figure~\ref{fig2}(c) and (d) shows the fitted $\mathrm{\rho_{xy}}$ curves for bulk and device MST, respectively. The red solid line shows the ordinary and anomalous contributions in $\mathrm{\rho_{xy}}$ with B $\|$ c at 5 K in both cases. The sharp peak region, in excess of the ordinary and anomalous Hall contribution, is attributed to the THE and is extracted by subtracting the $\mathrm{\rho_{xy}}$ curves by the red solid line data. The contour plots of the extracted THE amplitude for both the bulk and nanoflake device are shown in Fig.~\ref{fig2}(e),(f), respectively. From the figures, it is evident that the THE amplitude for the bulk persists up to a higher temperature when compared to the nanoflake device. This could be because of the COC state being more vulnerable to thermal fluctuations when one approaches nanoscale thickness limit compared to the bulk. The THE amplitude mimicking the sensitivity of the COC state to sample thickness suggests that the former might be governed by the latter.
\begin{figure}
  \includegraphics[width=\linewidth]{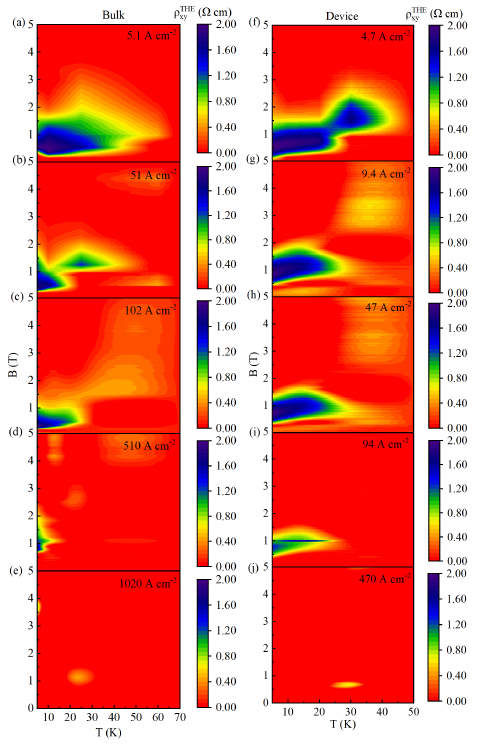}
  \caption{\textcolor{blue}{Contour plots of the THE amplitude ($\mathrm{\rho_{xy}^{THE}}$) for (a) - (e) Bulk and (f) - (j) Device, of MST, respectively, as a function of temperature and magnetic field for different applied currents.}}
  \label{fig3}
\end{figure}

To ascertain the causal connection of the COC and the observed THE in MST, we perform the magnetic field dependence of the Hall resistivity as a function of the applied current. Since it is well known that the strength of the COC decreases with increasing current \cite{Zhang2024CurrentSensitiveHall}, the relation of $\mathrm{\rho_{xy}^{THE}}$ to the current will be crucial in studying its actual origin. Figure~\ref{fig3} depicts the extracted THE amplitudes as a function of temperature and magnetic field for different applied currents for both bulk [Fig~\ref{fig3}(a) - (e)] and device [Fig~\ref{fig3}(f) - (j)], respectively. \textcolor{blue}{As the measurements of bulk and device were performed at current values with orders of magnitude difference, we employ current densities for meaningful comparison.} It is evident from the figure that upon increasing the current, the THE amplitudes decrease for both the bulk and the nanoflake device. \textcolor{blue}{This confirms that the THE signal is not an artifact of polaronic effect. Polaronic effects typically strengthen with increased carrier activation or Joule heating, i.e., with increasing current. Here, we see a complete opposite trend confirming the connection between THE and COC.} 

\textcolor{blue}{For bulk MST, the THE amplitude remains non-zero even up to 70 K when the applied current density is small (5.1 A cm$^{-2}$). As the current density increases to 51 A cm$^{-2}$, the region of the non-zero THE amplitude seems to split into two regimes of topological Hall signals [Fig.~\ref{fig3}(b)], which shrink and eventually vanishe upon increasing the current density to 1020 A cm$^{-2}$ as seen in Fig.~\ref{fig3}(e), similar to that reported in \cite{Zhang2024CurrentSensitiveHall}. This is expected as the COC is stronger for lower currents (leading to a lower current density) but on increasing the current (or current density) it starts to weaken. At high current density, the COC state ($\mathrm{\Psi_C}$) is destroyed completely, making way for a trivial state ($\mathrm{\Psi_T}$). Similar trend is observed to be followed by the THE signal of the bulk MST crystals, indicating that the two might be interconnected.} 

\textcolor{blue}{Although the THE signal from the nanoflake device of MST replicates the current dependence in case of the bulk MST, the device shows the disappearance of the THE behaviour at a much lower current density than the bulk, as is evident from, Fig.~\ref{fig3}(f) - (j). This reflects the reduced robustness of the COC state under reduced dimensionality. The lower critical current density for the disappearance of the THE contribution in the nanoflake device originates from the increased vulnerability of the COC state under reduced dimensionality. Thickness reduction limits the coherence volume of orbital currents, enhances surface-induced orbital dephasing, increases sensitivity to local electric fields and thermal fluctuations, and weakens interlayer phase locking. As a result, the orbital-coherent state collapses at lower current density in nanoflake devices compared to bulk crystals, while preserving the same underlying physical mechanism. This systematic reduction of the critical current density with thickness is a hallmark of a fragile, coherence-based orbital state rather than a conventional transport effect.}

\begin{figure}
  \includegraphics[width=\linewidth]{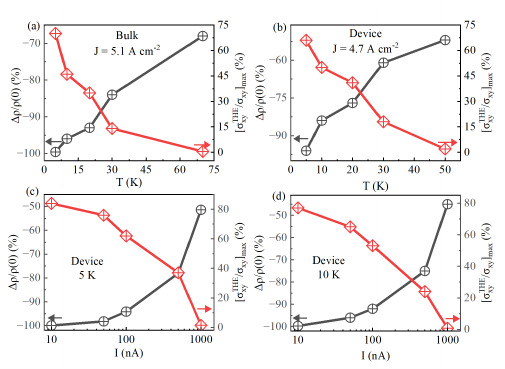}
  \caption{Temperature dependence of MR$\%$ and maximum relative topological Hall conductivity $\Big[\mathrm{\big[\sigma_{xy}^{THE}/\sigma_{xy}\big]_{max}} (\%)\Big]$ for (a) Bulk and (b) Device of MST, respectively. Current dependence of MR$\%$ and relative topological Hall conductivity $\Big[\mathrm{\big[\sigma_{xy}^{THE}/\sigma_{xy}\big]_{max}} (\%)\Big]$ for the device of MST at (c) 5 K and (d) 10 K, respectively. Arrows indicate the scale followed by the curves, respectively.}
  \label{fig4}
\end{figure}

From previous reports, it is well-established that the CMR portrayed by the MST is governed primarily by the COC \cite{Zhang2022ChiralOrbitalCurrents, Zhang2024CurrentSensitiveHall, PhysRevB.111.174419}. Thus, drawing a correlation between the THE and the magnetoresistance of MST in bulk and device, respectively, will provide a firm assurance of the COC origin of the THE. Figure~\ref{fig4}(a),(b) shows the temperature dependence of MR$\%$ and relative topological Hall conductivity $\Big[\mathrm{\big[\sigma_{xy}^{THE}/\sigma_{xy}\big]_{max}} (\%)\Big]$ for the bulk and device of MST, respectively. The MR$\%$ is calculated as, $\text{MR}\% = \mathrm{\frac{\Delta\rho}{\rho(0)} \times 100}$, where, $\mathrm{\frac{\Delta\rho}{\rho(0)} = \frac{\rho(B) - \rho(0)}{\rho(0)}}$ with $\rho(\mathrm{B})$ being the magnetoresistivity measured at an applied magnetic field of B T, and $\rho(0)$ the zero-field (ZF) resistivity. The hall conductivity is given by $\mathrm{\sigma_{xy} = \frac{\rho_{xy}}{\rho_{xx}^2 + \rho_{xy}^2}}$, where $\mathrm{\rho_{xx}}$ and $\mathrm{\rho_{xy}}$ are the longitudinal and Hall resistivity, respectively. Similarly, the topological Hall conductivity is given by $\mathrm{\sigma_{xy}^{THE} = \frac{\rho_{xy}^{THE}}{\rho_{xx}^2 + \rho_{xy}^2}}$, where $\mathrm{\rho_{xy}^{THE}}$ is the extracted topological Hall resistivity. 

As the temperature increases, spin-disorder scattering due to enhanced thermal fluctuation cannot be suppressed even with a higher magnetic field. This weakens the COC flowing along the Te edges and reduces the M$\mathrm{_{COC}}$. The CMR, promoted by the unique SOC arising from the coupling between M$\mathrm{_{COC}}$ and magnetic moments of Mn, M$\mathrm{_{Mn}}$, hence decreases with increasing temperature, as is evident from the black curve in Fig.~\ref{fig4}(a) and (b). The $\mathrm{\big[\sigma_{xy}^{THE}/\sigma_{xy}\big]_{max}} (\%)$ [red curve in Fig.~\ref{fig4}(a) and (b)] is seen to correlate well with the MR$\%$ for both the bulk and nanoflake device. Additionally, $\mathrm{\big[\sigma_{xy}^{THE}/\sigma_{xy}\big]_{max}} (\%)$ also decreases with temperature, eventually disappearing at higher temperatures, where the COC also gets completely suppressed. However, for the nanoflake device, the suppression of both [CMR and $\mathrm{\big[\sigma_{xy}^{THE}/\sigma_{xy}\big]_{max}} (\%)$] is much more rapid, highlighting the fragility of the COC state in lower dimensions.

As COC is also very sensitive to the applied current, the current dependence of MR$\%$ and relative topological Hall conductivity $\Big[\mathrm{\big[\sigma_{xy}^{THE}/\sigma_{xy}\big]_{max}} (\%)\Big]$ is also depicted in Fig.~\ref{fig4}(c) and (d) for the nanoflake device of MST at 5 K and 10 K, respectively. As the current is increased, the ferrimagnetic state as well as the COC state get suppressed. The zero field resistivity also decreases with increasing current which lead to a decrease in CMR with increased current as is evident from the black curves in Fig.~\ref{fig4}(c) and (d). The $\mathrm{\big[\sigma_{xy}^{THE}/\sigma_{xy}\big]_{max}} (\%)$] also decreases with increasing current [red curves in Fig.~\ref{fig4}(c) and (d)], indicating a similar origin as the CMR. At elevated temperatures, the suppression occurs at a lower current, consistent with reduced magnetic anisotropy and enhanced thermal fluctuations.

Thus, we observe that the THE in both bulk and nanoscale device of MST has a very strong correlation with the phenomena of CMR. As the latter is directly influenced by the COC in MST, the former could possibly have the same origin. It has already been reported that the COC develop a moment M$\mathrm{_{COC}}$ which induces a magnetic field B$_\mathrm{c}$ in real space, directed along the c-axis. This gives rise to a non-zero local scalar spin chirality, and the moment M$\mathrm{_{COC}}$ acts as the topological orbital moment (TOM). This couples with the local spins of Mn giving rise to spin-chiral interaction (SCI), mediated by SOC. Also, each individual COC can interact with each other, giving rise to a chiral-chiral interaction (CCI) \cite{Grytsiuk2020}. The formation of these two topological chiral interactions give rise to a finite spin chirality. Now, the THE contribution to the resistivity comes from two terms: $\mathrm{\rho_{xy}^{THE}} = \mathrm{\rho_c} + \mathrm{\rho_a}$, where $\mathrm{\rho_c}$ refers to the transverse charge current due to chiral orbital textures and $\mathrm{\rho_a}$ comes from chiral spin textures  \cite{PhysRevB.98.195439}. Hence, systems with strong dominating chiral orbital textures, like COC, can also show THE, which might be the reason behind the observed THE in MST.   

{\it Conclusion:---}
In this letter, we establish a direct correspondence between the COC state and the THE in MST. Our measurements on bulk and nanoflake devices reveal that the THE amplitude follows the evolution of the COC state with temperature and current, and exhibits a strong correlation with the CMR phenomena. These findings demonstrate that orbital textures can also act as a source of emergent gauge fields for conduction electrons, giving rise to topological transport responses without invoking complex spin textures, like skyrmions. Together with the framework of topological chiral interaction, our work highlights the active role of orbital magnetism in stabilizing chiral transport phenomena and points toward orbital engineering in 2D magnets as a powerful strategy for realizing dissipationless electronic states.

{\it Acknowledgements:---}
The authors acknowledge IIT Kanpur and the Department of Science and Technology, India, [Order No. DST/NM/TUE/QM-06/2019 (G)] for financial support. A.D. thanks PMRF for financial support.


\bibliography{cite} 


\end{document}